\documentclass[aps,prl,superscriptaddress,showpacs,reprint]{revtex4-1} 

\pdfoutput=1 

\usepackage{graphicx}
\usepackage{dcolumn}\newcolumntype{d}{D{.}{.}{-1}}
\usepackage{amssymb}
\usepackage{bm}
\usepackage{color}

\renewcommand{\vec}[1]{\ensuremath{\bm{#1}}}
\newcommand{\mat}[1]{\ensuremath{\bm{#1}}}

\newcommand{\CSC}{$^{13}$C~$\delta$} 
\newcommand{\CSH}{$^{1}$H~$\delta$} 
\newcommand{\CIEC}{1s~C~$\delta$} 
\newcommand{\ForceC}{$F_{\mathrm{C}}$}
\newcommand{\ForceH}{$F_{\mathrm{H}}$}

\newcommand{\mH}{mE$_h$}
\newcommand{\mHpB}{mE$_h$/$a_0$}

\usepackage[colorlinks=true,linkcolor=black,citecolor=black,filecolor=black,urlcolor=black]{hyperref}

\begin{document}


\title{Machine Learning for Quantum Mechanical Properties of Atoms in Molecules} 

\author{Matthias Rupp}
\email{mrupp@mrupp.info}
\altaffiliation{Current address: Fritz Haber Institute of the Max Planck Society, Faradayweg 4--6, 14195 Berlin, Germany.}
\author{Raghunathan Ramakrishnan}%
\author{O.~Anatole von Lilienfeld}
\email{anatole.vonlilienfeld@unibas.ch}
\affiliation{Institute of Physical Chemistry and National Center for Computational Design and Discovery of Novel Materials (MARVEL), Department of Chemistry, University of Basel, Klingelbergstr.~80, CH-4056 Basel, Switzerland}

\date{\today}

\begin{abstract}
We introduce machine learning models of quantum mechanical observables of atoms in molecules.
Instant out-of-sample predictions for proton and carbon nuclear chemical shifts, atomic core level excitations, and forces on atoms reach accuracies on par with density functional theory reference.
Locality is exploited within non-linear regression via local atom-centered coordinate systems.
The approach is validated on a diverse set of 9\,k small organic molecules.
Linear scaling of computational cost in system size is demonstrated for saturated polymers with up to sub-mesoscale lengths.
\end{abstract}

\pacs{03.65.-w,31.15.A-,31.15.E-,02.60.Ed}

\maketitle

This work has subsequently been published in the Journal of Physical Chemistry Letters 6(16): 3309--3313, American Chemical Society, 2015.
To access the final edited and published work see \href{http://dx.doi.org/10.1021/acs.jpclett.5b01456}{DOI 10.1021/acs.jpclett.5b01456}.

\bigskip


Accurate solutions to the many-electron problem in molecules have become possible due to progress in hardware and methods.~\cite{bc2010,dg2010,bta2009,bsrkl2014}
Their prohibitive computational cost, however, prevents both routine atomistic modeling of large systems and high-throughput screening.~\cite{hjo2000}
Machine learning (ML) models can be used to infer quantum mechanical (QM) expectation values of molecules, based on reference calculations across chemical space.~\cite{rtml2012,l2013}
Such models can speed up predictions by several orders of magnitude, demonstrated for relevant molecular properties such as enthalpies, entropies, polarizabilities, electron correlation, and, electronic excitations.~\cite{mrgvmhtmvl2013,rhtl2015,rdrl2015}

\begin{figure}
\includegraphics[width=\linewidth]{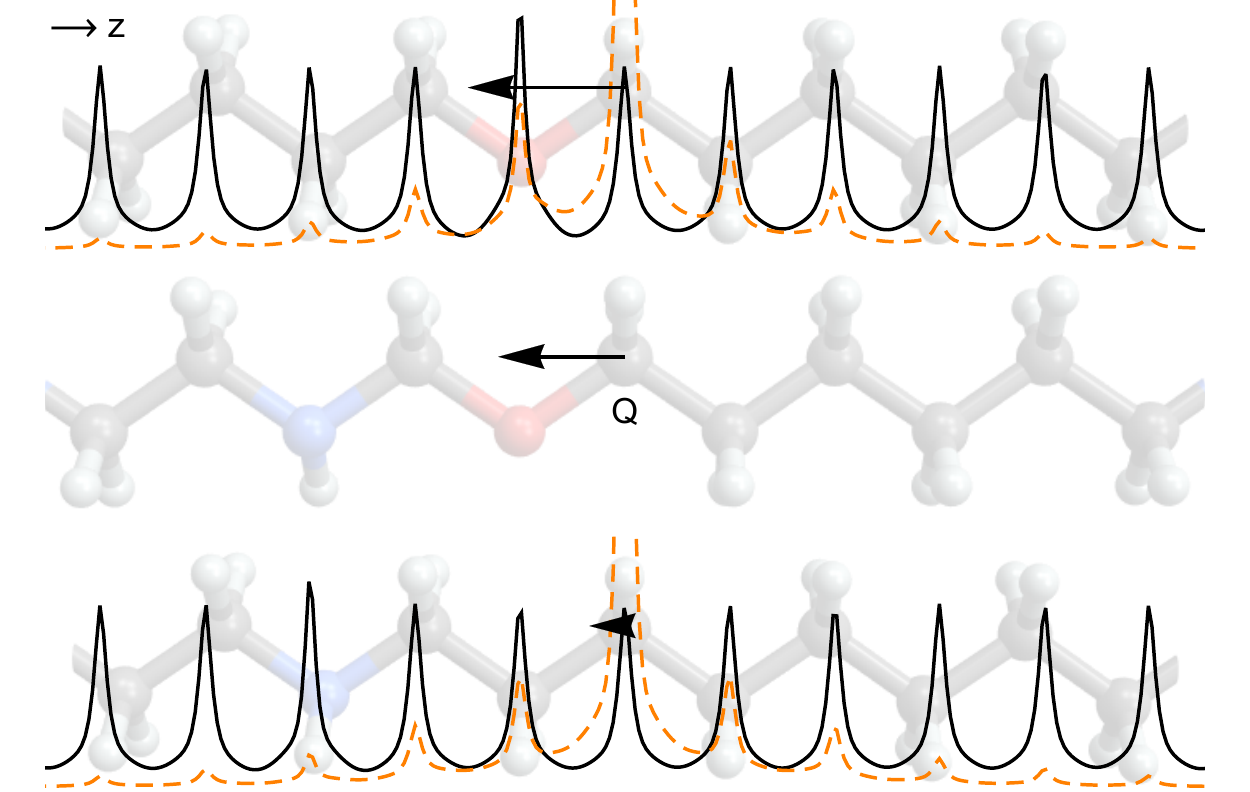}
\caption{Sketch illustrating local nature of atomic properties for the example of force $\langle \Psi | \partial_{\vec{R}_Q}\hat{H} | \Psi \rangle$ acting on a query atom
in a molecule (mid), inferred from similar atoms in training molecules (top, bottom).
Shown are force vectors (arrows), integrated electron density $\int \mathrm{d}x \, \mathrm{d}y \, n(\vec{r})$ (solid) and integrated electronic term of Hellmann-Feynman force along $z$ (dashed).\label{figSketch}}
\end{figure}

A major drawback is their lack of transferability, e.g., ML models trained on bond dissociation energies of small molecules will not be predictive for larger molecules.
In this work, we introduce ML models for properties of atoms in molecules. 
These models exploit locality to achieve transferability to larger systems \textit{and} across chemical space, for systems that are locally similar to the ones trained on (Fig.~\ref{figSketch}).
These aspects have only been treated in isolation before.\cite{rtml2012,lkv2015}

We model spectroscopically relevant observables, namely $^{13}$C and $^1$H nuclear magnetic resonance (NMR) chemical shifts \cite{fnShifts} and $1s$ core level ionization energies (CIE), as well as atomic forces, crucial for structural relaxation and molecular dynamics. 
Nuclear shifts and ionization energies are dominated by inherently local core electron-nucleus interactions. 
Atomic forces are expectation values of the differential operator applied to an atom's position in the Hamiltonian~\cite{f1939}, and scale quadratically with inverse distance.


Inductive modeling of QM properties of atoms in molecules constitutes a high-dimensional interpolation problem with spatial and compositional degrees of freedom.
QM reference calculations provide training examples $\{ (\vec{x}_i,y_i) \}_{i=1}^n$, where the $\vec{x}_i$ encode atoms in their molecular evironment and $y_i$ are atomic property values.
ML interpolation between training examples then provides predicted property values for new atoms.

The electronic Hamiltonian is determined by number of electrons, nuclear charges~$\{Z_I\}$ and positions~$\{\vec{R}_I\}$, which can be challenging for direct interpolation.~\cite{gvlds2015}
Proposed requirements for representations include uniqueness, continuity, as well as invariance to translation, rotation, and nuclear permutations.~\cite{lrrk2015}
For scalar properties (NMR, CIE), we use the sorted Coulomb matrix~\cite{rtml2012} to represent a query atom~$Q$ and its environment:
$M_{II} = 0.5 Z_I^{2.4}$ and $M_{IJ} = Z_I Z_J / |\vec{R}_I-\vec{R}_J|$, where atom indices $I,J$ run over $Q$ and all atoms in its environment, sorted by distance to~$Q$.
Note that all molecules in this study are neutral, and no explicit encoding of charge is necessary.

Atomic forces are vector quantities requiring a basis, which should depend only on the local environment; in particular, it should be independent of the global frame of reference used to construct the Hamiltonian in the QM calculation.
We project force vectors into a local coordinate system centered on atom~$Q$, and predict each component separately.
Later, the predicted force vector is reconstructed from these component-wise predictions.

We use principal component analysis (PCA) to obtain an atom-centered orthogonal three-dimensional local coordinate system.
In analogy to the electronic term in Hellmann-Feynman forces, 
$\int \mathrm{d} \vec{r} \,(\vec{r}-\vec{R}_Q) Z_Q\, n(\vec{r}) / \|\vec{r}-\vec{R}_Q\|^3$~\cite{f1939}, 
we weight atoms by $Z_I / \|\vec{R}_I-\vec{R}_Q\|^3$, increasing influence of heavy atoms and decreasing influence of distant atoms.
Non-degenerate PCA axes are unique only up to sign; we address this by defining the center of charge to be in the positive quadrant.
A matching matrix representation is obtained via $\mat{M}_{I} = (Z_I,X'_I,Y'_I,Z'_I)$, where $X',Y',Z'$ are projected atom coordinates, and rows are ordered by distance to central atom~$Q$, yielding an $m \times 4$ matrix, where $m$ is number of atoms.
In both representations, we impose locality by constraining $Q$'s environment to neighboring atoms within a sphere of radius~$\tau$.

For interpolation between atomic environments we use kernel ridge regression (KRR) \cite{htf2009}, a non-linear regularized regression method effectively carried out implicitly in a high-dimensional Hilbert space (``kernel trick'').~\cite{ss2002}
Predictions are linear combinations over all training examples in the basis of a symmetric positive definite kernel~$k$:
$f(\vec{z}) = \sum_{i=1}^n \alpha_i k(\vec{x}_i,\vec{z})$,
where $\vec{\alpha}$ are regression weights for each example, obtained from a closed-form expression minimizing the regularized error on the training data.
See Refs.~\cite{rtml2012,hmbfrsvltm2013,vslrckmb2015,lsphndrmb2015} for details.
As kernel~$k$, we use the Laplacian kernel 
$k(\vec{x},\vec{z}) = \exp \bigl( - \| \vec{x}-\vec{z} \|_1 / \sigma d \bigr)$,
where $\|\cdot\|_1$ is the $L^1$-norm, $\sigma$ is a length scale, and $d = \mathrm{dim}(\vec{x})$.
This kernel has shown best performance for prediction of molecular properties.~\cite{hmbfrsvltm2013}

\begin{figure}
	\includegraphics[width=\linewidth]{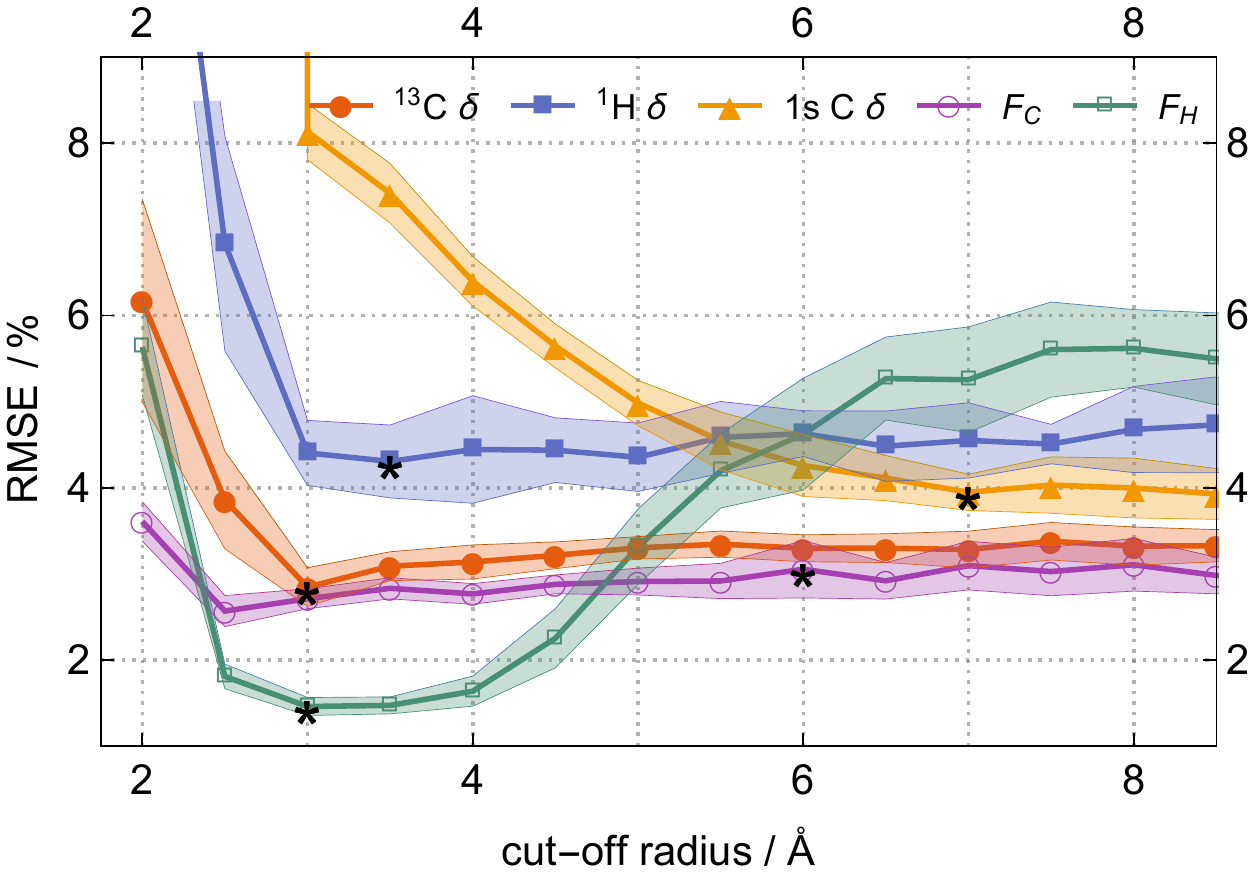}
\caption{Locality of properties, measured by model performance as a function of cut-off radius~$\tau$.
Root mean square error (RMSE) shown as fraction of corresponding property's range~\cite{fnPropertyRanges} for nuclear shifts (\CSC, \CSH), core level ionization energy (\CIEC), and atomic forces (\ForceC, \ForceH).
Asterisks~$\ast$ mark chosen values.
Shaded areas indicate 1.6 standard deviations over 15 repetitions.
\label{figTauConvergence}}
\end{figure}

Our models contain three free parameters: 
cut-off radius~$\tau$, regularization strength~$\lambda$, and kernel length scale~$\sigma$.
Regularization strength~$\lambda$, controlling the smoothness of the model, was set to a small constant ($10^{-10}$), forcing the model to fit the QM values closely.
As for length scales~$\sigma$, note that for the Laplacian kernel, non-trivial behavior requires $||\cdot,\cdot||_1 \approx \sigma$.
We set $\sigma$ to four times the median nearest neighbor $L^1$-norm distance in the training set.~\cite{md1989}
Cut-off radii~$\tau$ were then chosen to minimize RMSE in initial experiments (Fig.~\ref{figTauConvergence}).
For the comparatively insensitive \ForceC, other statistics (maxAE, $R^2$) yielded an unambiguous choice.

We used three datasets for validation:
For NMR chemical shifts and CIEs, both scalar properties, 
we employed a dataset of 9\,k synthetically accessible organic molecules containing 7--9 C, N, or O atoms, with open valencies saturated by H, a subset of a larger dataset.~\cite{rdbr2012,rdrl2014} 
Relaxation and property calculations were done at the DFT/PBE0/def2TZVP level of theory \cite{b2012,b2014c,pbe1996,wa2005,w2006f,ab1998} using {Gaussian} \cite{ftssrcsbmpetal2009}.
For forces, we distorted molecular equilibrium geometries using normal mode analysis~\cite{o1999,st1989,mp2012} by adding random perturbations in the range $[-0.2,0.2]$ to each normal mode, sampling homogeneously within an harmonic approximation.
Adding spatial degrees of freedom considerably increases the intrinsic dimensionality of the learning problem. 
To accommodate this, we reduced dataset variability to a subset of 168 constitutional isomers of C$_7$H$_{10}$O$_2$, with 100 perturbed geometries for each isomer.
Computationally inexpensive semi-empirical quantum chemistry approximations are readily available for forces.
We exploit this to improve accuracy by modeling the difference between baseline PM7~\cite{s1989} and DFT reference forces ($\Delta$-learning~\cite{rdrl2015}).
To demonstrate linear scaling of computational cost with system size, we used a third dataset of organic saturated polymers, namely linear polyethylene, the most common plastic, with random substitutions of some CH units with NH or O for chemical variety.
All prediction errors were measured on out-of-sample hold-out sets never used during training.

\begin{figure}
	\includegraphics[width=\linewidth]{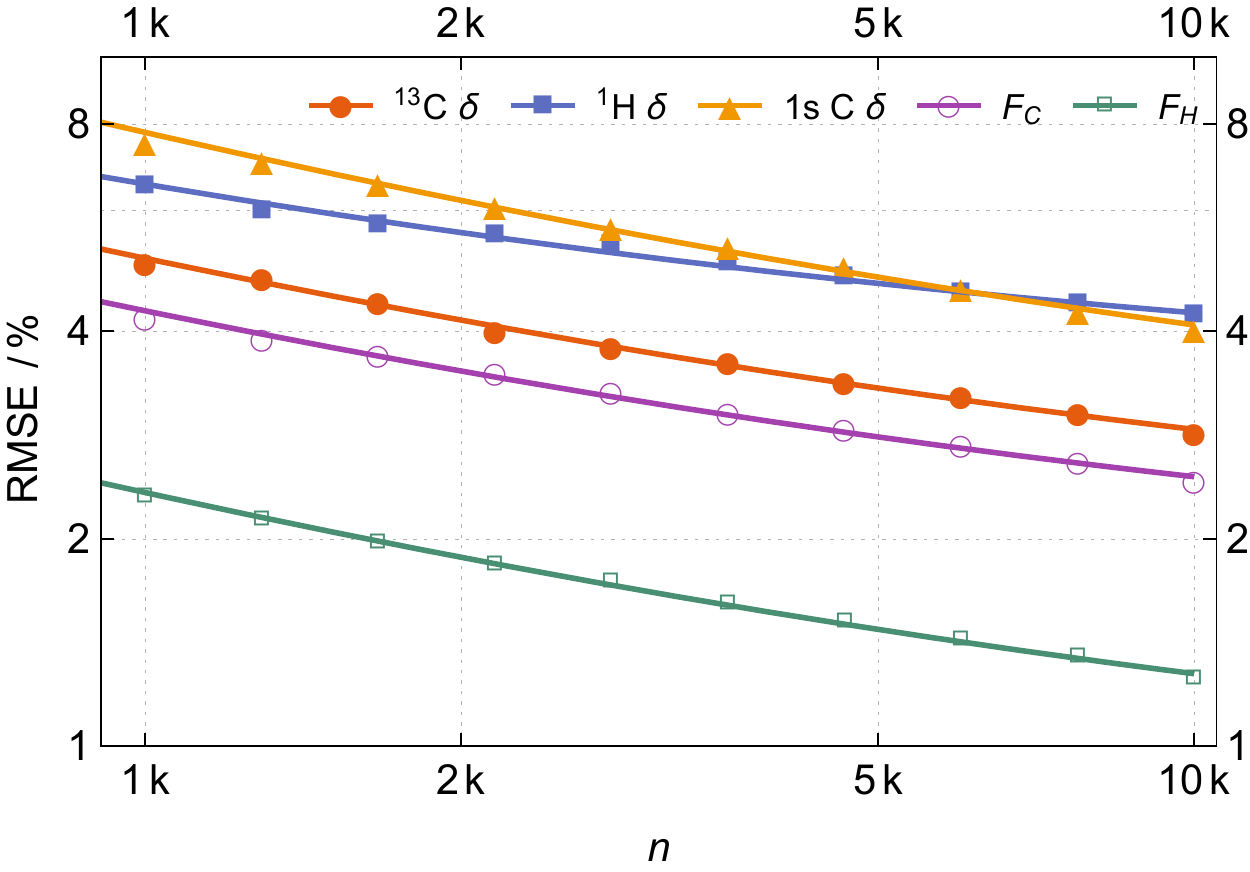}
\caption{Systematic improvement in accuracy of atomic property predictions with increasing training set size~$n$.
Root mean square error (RMSE) shown as fraction of corresponding property's range~\cite{fnPropertyRanges} for nuclear shifts (\CSC, \CSH), core level ionization energy (\CIEC), and atomic forces (\ForceC, \ForceH).
Values from 15 repetitions; see Table~\ref{tabResults} for ranges and standard deviations.
Solid lines are fits to theoretical asymptotic performance of $O(1/\sqrt{n})$.
\label{figLearningCurves}}
\end{figure}

Table~\ref{tabResults} presents performance estimates for models trained on 10\,k randomly chosen atoms, measured on a hold-out set of 1\,k other atoms.
Comparison with literature estimates of typical errors of the employed DFT reference method suggests in all cases that the ML models achieve similar accuracy---at negligible computational cost after training. 
Statistical learning theory shows that under certain assumptions the accuracy of a ML model asymptotically improves with increasing training set size as $O(1/\sqrt{n}$).~\cite{afs1992}
Fig.~\ref{figLearningCurves} presents corresponding learning curves for all properties.
Errors are shown as percentage of property ranges~\cite{fnPropertyRanges}, enabling comparison of properties with different units.
All errors start off in the single digit percent range at 1\,k training atoms, and decay systematically to roughly half their initial value at 10\,k training atoms. 

\begin{table*}[htbp]
\caption{\label{tabResults}%
Prediction errors for ML models trained on 10\,k atoms and predicting properties of 1\,k other out-of-sample atoms.
Calculated properties are NMR chemical shifts (\CSC, \CSH), core level ionization energy (\CIEC), and forces (\ForceC, \ForceH).$^a$
}
\small
\begin{tabular}{@{}llr@{ -- }l@{}r@{$\,\pm$}lr@{$\,\pm$}lr@{$\,\pm$}lr@{$\,\pm$}lr@{$\,\pm$}ld@{}}
	Property & \multicolumn{1}{@{}c@{}}{Ref.} & \multicolumn{2}{@{}c@{}}{Range} & \multicolumn{2}{@{}c@{}}{MAE} & \multicolumn{2}{@{}c@{}}{RMSE} & \multicolumn{2}{c}{maxAE} & \multicolumn{2}{c}{$R^2$} & \multicolumn{2}{c}{$\sigma$} & \multicolumn{1}{c}{$\tau$/\AA} \\ \hline
	\CSC/ppm        & 2.4  \cite{ab1998,dg2009,fmhbkto2014}        &    6 & 211 & 3.9  & 0.28 & 5.8  & 0.30 & 36  & 8.0 & 0.988 & 0.001 & 20   & 3.4 & 3   \\ 
	\CSH/ppm        & 0.11 \cite{fmhbkto2014,abcsd2006,hvh2015}    &    0 & 10  & 0.28 & 0.01 & 0.42 & 0.02 & 3.2 & 1.1 & 0.954 & 0.005 & 0.53 & 1.2 & 3.5 \\
	\CIEC/\mH       & 7.5  \cite{mbkst2002,kbswbs2002,hbst2011}    & -165 & -2  & 4.9  & 0.12 & 6.5  & 0.27 & 34  & 17  & 0.971 & 0.002 & 181  & 0.0 & 7   \\
	\ForceC/\mHpB   & 1    \cite{fnForceRefError}                  & -99  & 96  & 3.6  & 0.10 & 4.7  & 0.15 & 29  & 5.5 & 0.983 & 0.002 & 0.69 & 0.1 & 6   \\
	\ForceH/\mHpB   & 1    \cite{fnForceRefError}                  & -43  & 43  & 0.8  & 0.02 & 1.1  & 0.03 & 7.4 & 2.6 & 0.996 & 0.003 & 0.35 & 0.0 & 3   \\
\end{tabular}

\medskip\begin{minipage}{\linewidth}
$^a$ Shown are MAE of DFT reference from literature (Ref.), property ranges \cite{fnPropertyRanges}, mean absolute error (MAE), root mean squared error (RMSE), maximum absolute error (maxAE), squared correlation ($R^2$) and hyperparameters (kernel length scale~$\sigma$, cut-off radius~$\tau$).
Averages $\pm$ standard deviations over 15 randomly drawn training sets.
\end{minipage}
\end{table*}

\begin{figure}
\includegraphics[width=\linewidth]{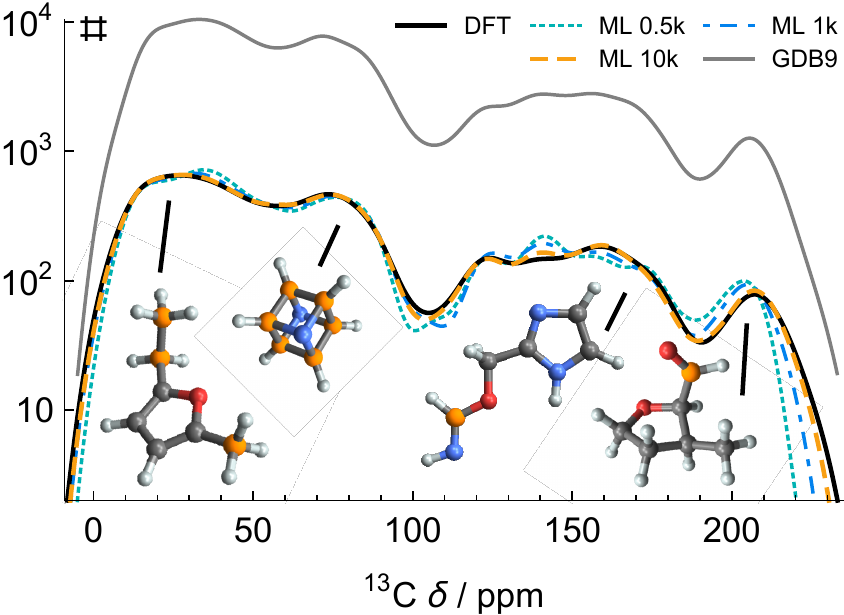}
\caption{\label{figPropertyDistributions}
Distribution of 50\,k $^{13}$C chemical shifts in 9\,k organic molecules.
ML predictions for increasing training set sizes approach DFT reference values.
Molecular structures highlight chemical diversity and effect of molecular environment on chemical shift of query atom (orange; see main text).
GDB9 corresponds to ML predictions for 847\,k carbon atoms in 134\,k similar molecules published in Ref.~\cite{rdrl2014}.
}
\end{figure}

ML predictions and DFT values for chemical shifts of all 50\,k carbon atoms in the dataset are featured in Fig.~\ref{figPropertyDistributions}.
The shielding of the nuclear spin from the magnetic field is strongly dependent on the atom's local chemical environment.
In accordance with the diversity of the dataset, we find a broad distribution with four pronounced peaks, characteristic of up- or downshifts of the resonant frequencies of nuclear carbon spin.
The peaks at 30, 70, 150, and 210\,ppm typically correspond to saturated sp$^3$-hybridized carbon atoms, strained sp$^3$-hybridized carbons, conjugated or sp$^2$-hybridized carbon atoms, and carbon atoms in carbonyl groups, respectively.
A ML model trained on only 500 atoms already reproduces all major peaks; larger training sets yield systematically improved distributions.
For 10\,k training examples predictions are hardly distinguishable from the DFT reference, except for a small deviation at 140\,ppm. 
Using the same model, we predicted shifts for 847\,k carbon atoms in all 134\,k molecules published in Ref.~\cite{rdrl2014}.
The resulting distribution is roughly similar, reflecting similar chemical composition of molecules in this much larger dataset, which is beyond the current limits of DFT reference calculations employed here.

The presented approach to model atomic properties scales linearly:
Since only a finite volume around an atom is considered, its numerical representation is of constant size; \footnote{Although the size of the representation may vary, it is bounded from above by a constant.}
in particular, it does not scale with the system's overall size.
Comparing atoms, and thus kernel evaluations, therefore requires constant computational effort, rendering the overall computational cost of predictions linear in system size, with small prefactor.
Furthermore, a form of chemical extrapolation can be achieved despite the fact that ML models are interpolation models. 
As long as local chemical environments of atoms are similar to those in the training set, the model can interpolate.
Consequently, using similar local ``building blocks'', large molecules can be constructed that are very different from the ones used in the training set, but amenable to prediction.

\begin{figure}
	\includegraphics[width=\linewidth]{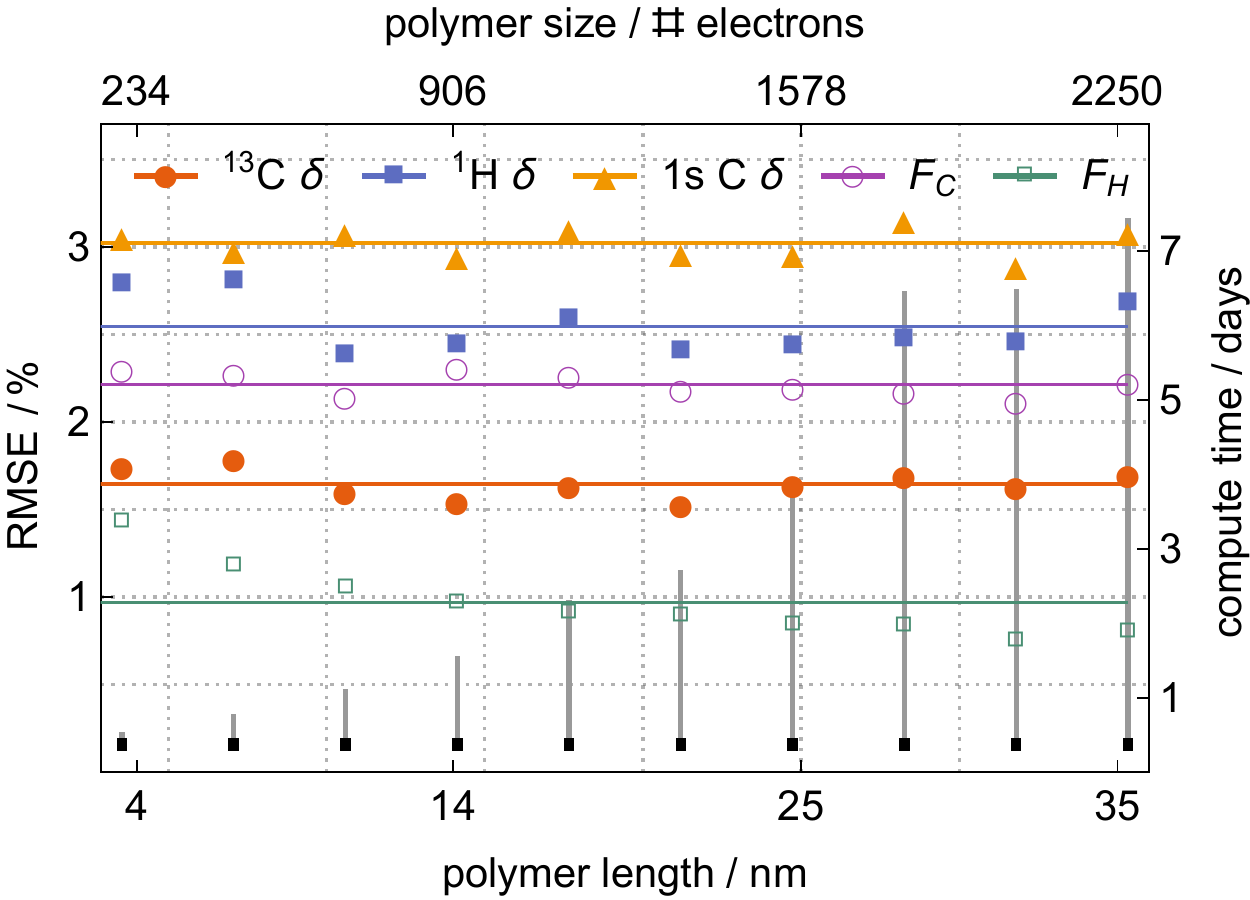}
\caption{Linear scaling and chemical extrapolation for ML predictions of saturated polymers of increasing length.
Shown are root mean square error (RMSE), given as fraction of corresponding property's range~\cite{fnPropertyRanges},
as well as indicative compute times of cubically scaling DFT calculations (gray bars) and ML predictions (black bars, enlarged for visibility), which scale linearly with low prefactor.
See Table~\ref{tabResults} for property ranges.
\label{figScaling}}
\end{figure}

To verify this, we trained a ML model on atoms drawn from the short polymers in the third dataset, then applied the same model to predict properties of atoms in polymers of increasing length.
Training set polymers had a backbone length of 29 C,N,O atoms; for validation, we used up to ten times longer backbones, reaching lengths of 355\,\AA{} and 696 atoms in total.
Fig.~\ref{figScaling} presents numerical evidence for excellent near-constant accuracy of model predictions, independent of system size, validated by DFT.
Although trained only on the smallest instances, the model's accuracy varies negligibly with system size, confirming both transferability and chemically extrapolative predictive power of the ML model.

Individual ML predictions are 4--5 orders of magnitude faster than reference DFT calculations.
Overall speed-up depends on dataset and reference method, and is dominated by training set generation, i.e., the ratio between number of predictions and training set size.
DFT and ML calculations were done on a high-performance compute cluster and a laptop, respectively.


In conclusion, we have introduced ML models for QM properties of atoms in molecules.
Performance and applicability have been demonstrated for chemical shifts, core level ionization energies, and atomic forces of 9\,k chemically diverse organic molecules and 168 isomers of C$_7$H$_{10}$O$_2$, respectively.
Accuracy of predictions is on par with the QM reference method.
We have used the ML model to predict chemical shifts of all 847\,k carbon atoms in the 134\,k molecules published in Ref.~\cite{rdrl2014}. 
Locality of modeled atomic properties is exploited through use of atomic environments as building blocks.
Consequently, the model scales linearly in system size, which we have demonstrated for saturated linear polymers over 30\,nm in length.
Results suggest that the model could be useful in mesoscale studies.

For the investigated molecules and properties the locality assumption, implemented as a finite cut-off radius in the representation, has proven sufficient.
This might not necessarily be true in general. 
The Hellmann-Feynman force, for example, depends directly on the electron density, which can be altered substantially due to long-range substituent effects such as those in conjugated $\pi$-bond systems. 
For other systems and properties, larger cut-offs or additional measures might be necessary. 

The presented ML models could also be used for nuclear shift assignment in NMR structure determination, for molecular dynamics of macro-molecules, or condensed molecular phases. 
We consider efficient sampling, i.e., improving the ratio of performance to training set size (``sample efficiency''), and improving representations to be primary challenges in further development of these models.

\begin{acknowledgments}
We thank Tristan Bereau, Zhenwei Li, and Kuang-Yu Samuel Chang for helpful discussions. 
OAvL acknowledges the Swiss National Science Foundation for support (SNF grant PP00P2\_138932).
Calculations were performed at sciCORE (\url{scicore.unibas.ch}) scientific computing core facility at University of Basel.
This research used resources of the Argonne Leadership Computing Facility at Argonne National Laboratory,
which is supported by the Office of Science of the U.S.~DOE under contract DE-AC02-06CH11357.
\end{acknowledgments}



\begin{thebibliography}{45}
\bibitem [{fnP()}]{fnPropertyRanges}%
  \BibitemOpen
  \href@noop {} {}\bibinfo {note} {Instead of $\max(p)-\min(p)$, which are
  extrema that strongly fluctuate across subsets of the data, we use the more
  robust 99\,\% and 1\,\% quantiles.}\BibitemShut {Stop}%
\bibitem [{fnF()}]{fnForceRefError}%
  \BibitemOpen
  \href@noop {} {}\bibinfo {note} {{Gaussian's} default convergence threshold
  of 1\,\mHpB{} was used as target accuracy.}\BibitemShut {Stop}%
\end{thebibliography}

\end{document}